\documentclass[12pt]{article}
\usepackage[all]{xy}

\usepackage{epic}
\usepackage{amsmath}[1996/11/01]
\usepackage{amssymb,amsthm,amsfonts,latexsym,epsfig,graphics}
 \def\beql#1#2\eeql{\begin{equation}\label{#1}#2\end{equation}}

\advance \textheight by -1 \topmargin
\advance \textheight by -1 \headheight
\advance \textheight by -1 \headsep
\advance \textheight by -1 \footskip
\vsize = \textheight
\hsize = \textwidth
\newcommand{\operp}{\mathop{\bigcirc\!\!\!\!\!\!\!\perp\,}} %orthog. dir. Summe
\DeclareMathOperator{\wt}{wt}
\DeclareMathOperator{\Char}{char}

\DeclareMathOperator{\diag}{diag}

\DeclareMathOperator{\GL}{GL}
\DeclareMathOperator{\Aut}{Aut}
\DeclareMathOperator{\Mon}{Mon}

\DeclareMathOperator{\id}{id}
\DeclareMathOperator{\SL}{SL}
\DeclareMathOperator{\PSL}{PSL}

\newtheorem{theorem}{Theorem}[section]

\newcommand{\bew}{\noindent\underline{Proof.}\ }
\newtheorem{rem}[theorem]{Remark}
\newtheorem{lemma}[theorem]{Lemma}
\newtheorem{proposition}[theorem]{Proposition}

\newtheorem{defn}[theorem]{Definition}

\newcommand{\Z}{{\mathbb{Z}}}

\newcommand{\F}{{\mathbb{F}}}
\newcommand{\N}{{\mathbb{N}}}

\newcommand{\eb}{\phantom{zzz}\hfill{$\square $}\smallskip}

\renewcommand{\em}{\sf}

\begin{document}
\begin{center}
{\Large {\bf On extremal self-dual ternary codes of length 48}}\\
\vspace{1.5\baselineskip}
{\em Gabriele Nebe} \\
\vspace*{1\baselineskip}
Lehrstuhl D f\"ur Mathematik, RWTH Aachen University\\
52056 Aachen, Germany \\
 nebe@math.rwth-aachen.de \\
\vspace{1.5\baselineskip}
\end{center}

{\sc Abstract.}
{\small All extremal ternary codes of length 48 that have some
automorphism of prime order $p\geq 5$ are equivalent to one of the
two known codes, the Pless code or the extended quadratic residue code.
}

\section{Introduction.}

The notion of an extremal code has been introduced in 
\cite{MallowsSloane}. 
As Andrew Gleason \cite{Gleason} 
remarks one may use  invariance properties 
of the weight enumerator of a self-dual code to deduce upper bounds on
the minimum distance. 
Extremal codes are self-dual codes that achieve these bounds.
The most wanted extremal code is a binary self-dual doubly even 
code of length 72 and minimum distance 16. 
One frequently used strategy is to classify extremal codes with 
a given automorphism, see \cite{Huffman} and \cite{CP} for the
first papers on this subject. 

Ternary codes have been studied in \cite{Huffmantern}. 
The minimum distance $d(C):=\min \{ \wt (c) \mid 0\neq c\in C \} $ of a
self-dual ternary code $C=C^{\perp } \leq \F_3^n$ of length $n$ is 
bounded by $$d(C) \leq 3 \lfloor \frac{n}{12} \rfloor +3 .$$
Codes achieving equality are called {\em extremal}. 
Of particular interest are extremal ternary codes of length a multiple of 12.
There exists a unique extremal code of length 12 (the extended ternary Golay code), 
two extremal codes of length 24 (the extended quadratic residue code $Q_{24} := \tilde{QR}(23,3)$ and the
Pless code $P_{24}$). For length 36, the Pless code yields one example of 
an extremal code. \cite{Huffmantern} shows that 
this is the only code with an automorphism of prime order $p\geq 5$, 
 a complete classification is yet unknown. 
The present paper investigates the extremal codes of length 48. 
There are two such codes known, the extended quadratic residue code $Q_{48}$ and the
Pless code $P_{48}$.
The computer calculations described in this paper show that these two codes are the only 
extremal ternary codes $C$ of length 48 for which the order of the
automorphism group is divisible by some prime $p\geq 5$. 
Theoretical arguments exclude all types of automorphisms that do not occur for the
two known examples.

\section{Automorphisms of codes.} 

Let $\F $ be some finite field, $\F^*$ its multiplicative group.
For any monomial transformation $\sigma \in \Mon_n(\F ) := \F^* \wr S_n $,
 the image $\pi (\sigma )\in S_n$ is called the 
{\em permutational part} of $\sigma $.
Then $\sigma $ has a unique expression as 
$$\sigma = \diag (\alpha _1 ,\ldots , \alpha _n) \pi (\sigma ) = m(\sigma ) \pi (\sigma ) $$
and $m(\sigma )$ is called the {\em monomial part} of $\sigma $.
For a code $C\leq \F^n$ we let 
$$\Mon (C):= \{ \sigma \in \Mon _n(\F ) \mid \sigma (C) = C \} $$
be the full monomial automorphism group of $C$.

We call a code $C\leq \F^n$ an orthogonal direct sum, if there are codes 
$C_i \leq \F^{n_i}$ ($1\leq i \leq s > 1$) of length $n_i$ such that 
$$C \sim \operp _{i=1} ^s C_i = \{ (c^{(1)}_1,\ldots , c^{(1)}_{n_1}, \ldots 
, c^{(s)}_1,\ldots , c^{(s)} _{n_s }  ) \mid c^{(i)} \in C_i (1\leq i \leq s) \} .$$

\begin{lemma}\label{orth}
Let $C\leq \F^n$ be not an orthogonal direct sum. 
Then the kernel of the restriction of $\pi $ to $\Mon (C)$ is isomorphic to $\F^* $.
\end{lemma}

\bew
Clearly $\F^* C = C $ since $C$ is an $\F$-subspace. 
Assume that $\sigma := \diag ( \alpha _1 ,\ldots , \alpha _n)  \in \Mon (C) $ with 
$\alpha _i \in \F^*$, not all equal. 
Let $\{ \alpha _1,\ldots , \alpha _n \} = \{ \beta _1,\ldots , \beta _s \}$ 
with pairwise distinct $\beta _i$.
Then 
$$C = \operp _{i=1}^s \ker (\sigma - \beta _i \id) $$ 
is the direct sum of eigenspaces of $\sigma $. Moreover the 
standard basis is a basis of eigenvectors of $\sigma $ so this is an orthogonal direct sum. 
\eb

In the investigation of possible automorphisms of codes, the following strategy 
has proved to be very fruitful (\cite{Huffman}, \cite{XXX}). 

\begin{defn}
Let $\sigma \in \Mon (C) $ be an automorphism of $C$. 
Then $\pi (\sigma ) \in S_n $ is a direct product of disjoint cycles of lengths 
dividing the order of $\sigma $. 
In particular if the order of $\sigma $ is some prime $p$, then 
we say that $\sigma $ has cycle type $(t,f)$, if 
$\pi (\sigma )$ has $t$ cycles of length $p$ and $f$ fixed points 
(so $pt+f = n$). 
\end{defn}

\begin{lemma} \label{nota}
Let $\sigma \in \Mon(C) $ have prime order $p$.
\\
(a) If $p$ does not divide $|\F^*|$ then
 there is some element $\tau \in \Mon _n(\F)  $ such that 
$m(\tau \sigma \tau^{-1}) = \id $.
Replacing $C$ by $\tau(C) $ we hence may
assume that $m(\sigma ) = 1 $.
\\
(b) Assume that $p$ does not divide  $\Char (\F )$,
 $m(\sigma ) = 1$, and $\pi (\sigma ) = (1,\ldots, p) \cdots ((t-1)p+1,\ldots , tp) 
(tp+1) \cdots (n) $. 
Then $C = C(\sigma ) \oplus E$, where 
$$C(\sigma ) = \{ c\in C \mid c_1=\ldots = c_p, c_{p+1} = \ldots = c_{2p}, \ldots 
c_{(t-1)p+1} = \ldots = c_{tp} \} $$ is the fixed code of $\sigma $ and 
$$E = \{ c\in C \mid \sum _{i=1}^p c_i = \sum _{i=p+1}^{2p} c_i = \ldots = \sum _{i=(t-1)p+1}^{tp} c_i =c_{tp+1} = \ldots = c_{n} = 0 \} $$
is the unique $\sigma $-invariant  complement of $C(\sigma ) $ in $C$.
\\
(c) Define two projections 
$$\begin{array}{lll} \pi _t: & C(\sigma ) \to \F ^t, & \pi _t(c):= (c_p,c_{2p},\ldots , c_{tp})  \\ 
 \pi _f: & C(\sigma ) \to \F ^f, & \pi _f(c):= (c_{tp+1},c_{tp+2}\ldots , c_{tp+f})  \end{array} $$ 
So $C(\sigma ) \cong (\pi_t(C(\sigma )), \pi _f(C(\sigma )) =: C(\sigma)^* $.
 If $C= C^{\perp }$ is self-dual with respect to $(x,y):=\sum _{i=1}^n x_i\overline{y_i} $,
then $C(\sigma )^* \leq \F ^{t+f}$ 
is a self-dual code with respect to the inner product 
$(x,y) := \sum _{i=1}^t px_i\overline{y_i} + \sum _{j=t+1}^{t+f} x_j \overline{y_j} $. 
\\
(d) In particular $\dim (C(\sigma )) = (t+f)/2$ and 
 $\dim (E) = t(p-1)/2$.
\end{lemma}

\bew
Part (a) follows from the Schur-Zassenhaus theorem in finite group theory. 
For the ternary case see \cite[Lemma 1]{Huffmantern}.
\\
(b) and (c) are similar to \cite[Lemma 2]{Huffman}. 
\eb

In the following we will keep the notation of the previous lemma and 
regard the fixed code $C(\sigma )$. 

\begin{rem}\label{lendC}
If $f\leq d(C) $ then $t\geq f$.
\end{rem} 

\bew
Otherwise the kernel  $K:=\ker (\pi _t) = \{ (0,\ldots , 0, c_1,\ldots ,c_f) \in C(\sigma )\} $
is a nontrivial subcode of minimum distance  $\leq f < d(C)$.
\eb

The way to analyse the code $E$ from Lemma \ref{nota} is based on the 
following remark.

\begin{rem}
Let $p \neq \Char(\F) $ be some prime and $\sigma \in \Mon _n(\F) $ be an
element of order $p$. Let 
$$X^p-1 = (X-1) g_1 \ldots g_m \in \F [X ] $$ 
be the factorization of $X^p-1$ into irreducible polynomials. 
Then all factors $g_i$ have the same degree $d  = |\langle |\F | +p \Z \rangle |$,
the order of $|\F |$ mod $p$.
\\
There are polynomials $a_i \in \F[X ]$ ($0\leq i \leq m$) such that 
$$1 = a_0  g_1 \ldots g_m + (X-1) \sum _{i=1}^m a_i \prod _{j\neq i} g_j .$$
Then the primitive idempotents in $\F[X]/(X^p-1)$ are given by the classes of 
$$\tilde{e}_0 = a_0  g_1 \ldots g_m, \tilde{e}_i = a_i \prod _{j\neq i} g_j (X-1) , 1\leq i \leq m .$$ 
Let $L$ be the extension field of $\F $ with $[L:\F ] = d$.
Then the group ring 
$$\F[X]/(X^p-1) = 
\F \langle \sigma \rangle \cong \F \oplus \underbrace{L\oplus \ldots \oplus L }_m $$
is a commutative semisimple $\F$-algebra. 
Any code $C\leq \F^n$ with an automorphism $\sigma \in \Mon (C)$ is a
module for this algebra. 
Put $e_i := \tilde{e}_i (\sigma ) \in \F[\sigma ]$.
Then $C=Ce_0\oplus Ce_1\oplus \ldots \oplus Ce_m $ with $Ce_0 = C(\sigma ) $, 
$E = Ce_1\oplus \ldots \oplus Ce_m $.
Omitting the coordinates of $E$ that correspond to the fixed points of $\sigma $, 
the codes $Ce_i$ are $L$-linear codes of length $t$. 
\\
Clearly $\dim _{\F }(E) = d \sum _{i=1}^m \dim _L(Ce_i) $.
\\
If $C$ is self-dual then $\dim (E) = t \frac{p-1}{2} $.
\end{rem}

\section{Extremal ternary codes of length 48.} 

Let $C=C^{\perp } \leq \F_3^{48}$ be an extremal self-dual ternary code of length 48,
so $d(C) = 15$.

\subsection{Large primes.} 

In this section we prove the main result of this paper.

\begin{theorem}
Let $C= C^{\perp } \leq \F_3^{48}$ be an extremal self-dual code 
with an automorphism of prime order $p \geq 5$. 
Then $C$ is one of the two known codes.
So either $C=Q_{48}$ is the extended quadratic residue code of length 48 
with automorphism group $$\Mon(C) = C_2 \times \PSL_2(47) \mbox{ of order } 2^5\cdot 3\cdot 23\cdot 47 $$
or $C=P_{48}$ is the Pless code with automorphism group 
$$\Mon (C) = C_2\times \SL _2(23) . 2 \mbox{ of order } 2^6 \cdot 3 \cdot 11 \cdot 23 .$$
\end{theorem} 

\begin{lemma}
Let $\sigma \in \Mon(C) $ be an automorphism of prime order $p \geq 5$. 
Then either $p=47$ and $(t,f) = (1,1)$ or
$p=23$ and $(t,f) = (2,2)$ or $p=11$ and $(t,f) = (4,4) $. 
\end{lemma}

\bew
For the proof we use the notation of Lemma \ref{nota}.
In particular we let $K:=\ker(\pi _t) =\{ (0,\ldots , 0 , c_1,\ldots ,c_f) \in C(\sigma )\}$ 
and put $K^*:= \{ (c_1,\ldots , c_f) \mid (0,\ldots , 0 , c_1,\ldots ,c_f) \in C(\sigma )\}$.
Then 
$$K^* \leq \F_3^f ,\ d(K^*) \geq 15 ,\ 
\dim (K^*) \geq \frac{f-t}{2}  .$$
Moreover $tp+f = 48$. 
\\
{\bf 1) If $t=1$ then $p=47$.} \\
If $p=47$, then $t=f=1$. 
\\
So assume that $p<47$ and $t=1$. 
Then the code $E$ has length $p$ and dimension $(p-1)/2$, therefore $p\geq d(C) = 15$.
So $p\geq 17$ and $f\leq 48-17 = 31$.
\\
Then $K^*\leq \F_3^{f}$ has dimension $(f-1)/2$ 
and minimum distance $d(K^*) \geq 15$. 
From the bounds given in \cite{codetables} there is no such possibility for 
$f\leq 31 $. 
\\
{\bf 2) If $t=2$ then $p=23$}. \\
Assume that $t=2$.
Since $2\cdot p \leq 48$ we get $p\leq 23$ and if $p=23$, then $(t,f) = (2,2)$.
\\
So assume that $p<23$. The code $E$ is a non-zero code of length $2p$ and minimum
distance $\geq 15$, so $2p \geq 15$ and  $p$ is one of $11, 13, 17, 19$, 
and $f= 26, 22, 14, 10 $.
The code $K^* \leq \F_3^f$ has
dimension  $\geq f/2 -1 $
and minimum distance $\geq 15$. Again by \cite{codetables} there is no such code.
\\
{\bf 3) $p\neq 13 $}. \\
For $p=13$ one now only has the possibility $t=3$ and $f=9$. 
The same argument as above constructs a code $K^*\leq \F_3^9$ 
of dimension at least $(f+t)/2-t=3$ of minimum distance $\geq 15 > f$ which is 
absurd. 
\\
{\bf 4) If $p=11$, then $t=f=4$.} \\ 
Otherwise $t=3$ and $f=15$ and the code $K^*$ as above has length 15, dimension 
$\geq 6$  and minimum
distance $\geq 15$ which is impossible. 
\\
{\bf 5) If $p=7$ then $t=f=6$.} \\
Otherwise $t=3,4,5$ and $f=27,20,13$ and the code $K^*$ as above has dimension 
$\geq (f+t)/2-t = 12, 8, 4 $, length $f$, minimum distance $\geq 15$
 which is impossible by \cite{codetables}.
\\
{\bf 6) $p\neq 7$.} \\
Assume that $p=7$, then $t=f=6$ and the kernel $K$ of the projection of
$C(\sigma )$ onto the first $42$ components is trivial. 
So the image of the projection is $\F_3^6\otimes \langle (1,1,1,1,1,1,1) \rangle $,
 in particular it  contains the 
vector $(1^7,0^{35}) $ of weight 7. 
So $C(\sigma ) $ contains some word $(1^7,0^{35},a_1,\ldots , a_6) $ of weight $\leq 13$ 
which is a contradiction.
\\
{\bf 7) If $p=5$ then $t=f=8$ or $t=9$ and $f=3$.} \\
Otherwise $t=3,4,5,6,7$ and $f=33,28,23,18,13$ and the code $K^*\leq \F_3^f$ has dimension 
$\geq (f+t)/2-t = 15, 12,9,6,3$ and minimum distance $\geq 15$ 
 which is impossible by \cite{codetables}.
\\
{\bf 8) $p\neq 5$.} \\
Assume that $p=5$. Then either $t=8$ and the projection of $C(\sigma )$ onto
the first $8\cdot 5$ coordinates is $\F_3^8 \otimes \langle (1,1,1,1,1) \rangle $ and
contains a word of weight 5. 
But then $C(\sigma )$ has a word of weight $w$ with $5<w\leq 5+8=13 $ a contradiction.
\\
The other possibility is $t=9$. Then the code $E = E^{\perp }$ is a Hermitian 
self-dual code of length 9 over the field with $3^4=81$ elements, 
which is impossible, since the length of such a code is 2 times the dimension and
hence even.
\eb

\begin{lemma}
If $p=11$ then $C\cong P_{48}$.
\end{lemma} 

\bew
Let $\sigma \in \Mon (C)$ be of order 11. 
Since
$(x^{11} -1) = (x-1) gh \in \F_3[x]$ for irreducible polynomials $g,h$ of degree 5, 
$$\F_3 \langle \sigma \rangle  \cong \F_3 \oplus \F_{3^{5}} \oplus \F_{3^{5}} .$$
Let $e_1,e_2,e_3 \in \F_3\langle \sigma \rangle $ denote the primitive idempotents. 
Then $C= Ce_1 \oplus Ce_2 \oplus Ce_3 $ with 
$C(\sigma ) = Ce_1 = Ce_1^{\perp }$ of dimension 4
and $Ce_2 =Ce_3^{\perp } \leq  (\F_{3^{5}} \oplus \F_{3^5})^4 $. 
Clearly the projection of $C(\sigma )$ onto the first $44$ coordinates is injective.
Since all weights of $C$ are multiples of 3 and $\geq 15$,
 this leaves just one possibility 
for $C(\sigma )$: 
$$G0 = (L0 | R0 ) :=\left( \begin{array}{cccc|cccc} 
1^{11} & 0^{11} & 0^{11} & 0^{11} & 1 & 1 & 1 & 1 \\
0^{11} & 1^{11} & 0^{11} & 0^{11} & 1 & 1 & -1 & -1 \\
0^{11} & 0^{11} & 1^{11} & 0^{11} & 1 & -1 & 1 & -1 \\
0^{11} & 0^{11} & 0^{11} & 1^{11} & 1 & -1 & -1 & 1 \end{array} \right) .$$
The cyclic code $Z$ of length 11 with generator polynomial $(x-1)g$ (and similarly 
the one with generator polynomial $(x-1)h$) has weight enumerator
$$x^{11}+132 x^5y^6  + 110 x^2 y^9 $$
in particular it contains more words of weight 6 than of weight 9. 
This shows that the dimension of $Ce_i$ over $\F_{3^5}$ is 2 for both $i=2,3$, 
since otherwise one of them has dimension $\geq 3$ and therefore contains all
words $(0,0,c,\alpha c)$ for all $c\in Z$ and some $\alpha \in \F_{3^5}$.
Not all of them can have weight $\geq 15$. 
Similarly one sees that the codes $Ce_i \leq \F_{3^5}^4$ have minimum 
distance 3 for $i=2,3$. 
So we may choose generator matrices 
$$G1:= \left( \begin{array}{cccc} 1 & 0 & a & b \\ 0 & 1 & c  & d \end{array} \right) ,\
G2:= \left( \begin{array}{cccc} 1 & 0 & a' & b' \\ 0 & 1 & c'  & d' \end{array} \right) $$
with $\left( \begin{array}{cc} a & b \\ c & d \end{array} \right) \in \GL_2(\F _{3^5}) $
and
$\left( \begin{array}{cc} a' & b '\\ c' & d' \end{array} \right) = 
-\left( \begin{array}{cc} a & b \\ c & d \end{array} \right)  ^{-tr}  $. 
To obtain $\F_3$-generator matrices for the corresponding codes 
$Ce_2$ and $Ce_3$ of length 48, we choose a generator matrix $g_1\in \F_3^{5\times 11}$ of the 
cyclic code $Z$ of length 11 with generator polynomial $(x-1)g$, and the corresponding 
dual basis $g_2\in \F_3^{5\times 11}$
 of the cyclic code with generator polynomial $(x-1)h$.
We compute the action of $\sigma $ (the multiplication with $x$) and represent
this as left multiplication with $z_{11} \in \F_3^{5\times 5} $ 
on the basis $g_1$. 
If $a = \sum _{i=0}^4 a_i z_{11}^i \in \F_{3^5} $ with $a_i \in \F_3$, then 
the entry $a$ in $G1$ is replaced by $\sum _{i=0}^4 a_i z_{11}^i g_1 \in \F_3^{5\times 11}$. 
Analogously for $G2$, where we use of course the matrix $g_2$ instead of $g_1$.
\\
Replacing the code by an equivalent one we may choose $a,b,c$ as orbit 
representatives of the action of $\langle - z_{11} \rangle $ on $\F_{3^5}^*$. 
\\
A generator matrix of $C$ is then given by 
$$ \left( \begin{array}{cc}  L0 & R0 \\ 
G1 & 0 \\ 
G2 & 0 \end{array} \right) .$$
All codes obtained this way are equivalent to the Pless code $P_{48}$.
\eb

\begin{lemma}
If $p=23$ then $C\cong P_{48}$ or $C\cong Q_{48}$.
\end{lemma} 

\bew
Let $\sigma \in \Mon (C)$ be of order 23. 
Since
$(x^{23} -1) = (x-1) gh \in \F_3[x]$ for irreducible polynomials $g,h$ of degree 11, 
$$\F_3 \langle \sigma \rangle  \cong \F_3 \oplus \F_{3^{11}} \oplus \F_{3^{11}} .$$
Let $e_1,e_2,e_3 \in \F_3\langle \sigma \rangle $ denote the primitive idempotents. 
Then $C= Ce_1 \oplus Ce_2 \oplus Ce_3 $ with 
$C(\sigma ) = Ce_1 = Ce_1^{\perp }$ of dimension 2
and $Ce_2 =Ce_3^{\perp } \leq ( \F_{3^{11}} \oplus \F_{3^{11}} )^2 $. 
Since all weights of $C$ are multiples of 3, this leaves just one possibility 
for $C(\sigma) $ (up to equivalence): 
$$C(\sigma ) = \langle (1^{23},0^{23},1,0) , (0^{23},1^{23},0,1) \rangle .$$
The codes $Ce_2$ and $Ce_3$ are codes of length 2 over $\F_{3^{11}}$ such that 
$\dim (Ce_2) + \dim (Ce_3) = 2$. 
Note that the alphabet $\F_{3^{11}}$ is identified with the cyclic code of length 23
with generator polynomial $(x-1)g$ resp. $(x-1)h$. 
These codes have minimum distance $9 < 15$, so 
$\dim (Ce_2) = \dim (Ce_3) = 1$ and both codes have a generator matrix of the form
$(1, t)$ (resp. $(1,-t^{-1})$) for $t \in \F_{3^{11}}^*$. 
Going through all possibilities for $t$ (up to the action of the 
subgroup of $\F_{3^{11}}^*$ of order 23) the only codes $C$ for which 
$C(\sigma ) \oplus Ce_2 \oplus Ce_3 $ have minimum distance $\geq 15 $ are
the two known extremal codes $P_{48}$ and $Q_{48}$.
\eb

\begin{lemma}
If $p=47$ then $C\cong Q_{48}$.
\end{lemma} 

\bew
The subcode $C_0 := \{ c \in \F_3^{47} \mid (c,0) \in C  \} $ is a cyclic code of length 
47, dimension 23 and minimum distance $\geq 15$.
Since $x^{47}-1 = (x-1) gh \in \F_3[x]$ for irreducible polynomials $g,h$ of degree 23, 
$C_0$ is the cyclic code with generator polynomial $(x-1) g$ (or equivalently $(x-1)h$) 
and  $C = \langle (C_0,0) , {\bf 1 } \rangle \leq \F_3^{48} $ is the extended quadratic residue code. 
\eb

\subsection{Automorphisms of order 2.} 

As above let $C=C^{\perp } \leq \F_3^{48} $ be an extremal self-dual ternary code. 
Assume that $\sigma \in \Mon (C)$ such that the permutational part
 $\pi (\sigma )$ has order 2.
Then $\sigma ^2 = \pm 1 $ because of Lemma \ref{orth}.
If $\sigma ^2 = -1 $, then $\sigma $ is conjugate to a block diagonal matrix 
with all blocks $\left(\begin{array}{cc} 0 & 1 \\ -1 & 0 \end{array} \right) =: J$
and $C$ is a Hermitian self-dual code of length 24 over $\F_9$. 
Such automorphisms $\sigma $ with $\sigma ^2=-1$ occur for both 
known extremal codes. 

If $\sigma ^2 = 1$, then $\sigma $ is conjugate to a block diagonal matrix
$$\sigma \sim \diag( \left(\begin{array}{cc} 0 & 1 \\ 1 & 0 \end{array} \right) ^t ,
 1^f , (-1)^a )$$
for $t,a,f\in \N_0 $, $2t+a+f = 48 $. 

\begin{proposition}\label{aut2}
Assume that $\sigma \in \Mon (C) $, $\sigma ^2=1$ and $\pi (\sigma ) \neq 1$. 
Then either $(t,a,f) = (24,0,0) $ or $(t,a,f) = (22,2,2)$. 
Automorphisms of both kinds are contained in $\Aut (P_{48})$.
\end{proposition}

\bew
{\bf 1) Wlog $f\leq a$.}  \\
Replacing $\sigma $ by $-\sigma $ we may assume without loss of generality that $f\leq a$. 
\\
{\bf 2) $f-t\in 4 \Z $.}  \\
By Lemma \ref{nota} the code $C(\sigma )^* \leq \F_3^{t+f}$ is a self-dual 
code with respect to the inner product $(x,y) = -\sum_{i=1}^t x_iy_i + \sum _{j=1}^f x_jy_j$.
This space only contains a self-dual code if  $f-t$ is a multiple of 4. 
\\
{\bf 3)  $t+f\in \{22, 24 \} .$} \\
The code $C(\sigma )^* \leq \F_3^{t+f} $ has dimension $\frac{t+f}{2}$ and 
minimum distance $\geq 15/2$ and hence minimum distance $\geq 8$.
 By \cite{codetables} this implies that $t+f \geq 22$.
Since $t+a \geq t+f $ and $(t+a) + (t+f) = 48$ this only leaves these two possibilities.
\\
{\bf 4) $t+f\neq 22$.}  \\
We first treat the case $f\leq 14$. Then $K^*\cong \ker(\pi _t)$ is a 
code of length $f\leq 14$ and minimum distance $\geq 15$ and hence trivial. 
So $\pi _t$ is injective and
 $$C(\sigma ) \cong D:=\pi _t(C(\sigma )) \leq \F_3^t, 
\dim (D) = 11 \mbox{, and } d(D) \geq \lceil \frac{15-f}{2} \rceil 
.$$ 
Using \cite{codetables} and the fact that $f-t$ is a multiple of $4$, 
this only leaves the cases $(t,f) \in \{ (19,3), (21,1) \} $.
To rule out these two cases we use the fact that 
$D$ is the dual of the self-orthogonal ternary code $D^{\perp} = \pi_t(\ker(\pi _f))$. 
The bounds in \cite{Annika} give 
$d(D) \leq 5< \frac{15-3}{2}$ for $t=19$ and $d(D) \leq 6 < \frac{15-1}{2}$ 
for $t=21$. 
\\
If $f\geq 15$, then $t\leq 7$ and $K^* \cong \ker (\pi _t)$ has dimension 
$f-t > 0$ and minimum distance $\geq 15$. This is easily ruled out by the known 
bounds (see \cite{codetables}).
\\
{\bf 5) If $t+f = 24$ then either $(t,f) = (24,0)$ or $(t,f) = (22,2)$.} 
\\
Again the case $f>t$ is easily ruled out using dimension and minimum distance of $K^*$
as before. 
So assume that $f\leq t $ and let $D = \pi _t(C(\sigma ) ) $ as before. 
Then $\dim (D) = 12$ and using \cite{codetables} one gets that 
$$(t,f) \in \{ (24,0), (22,2) , (20,4) \} .$$
Assume that $t=20$. Then there is some self-dual code $\Lambda = \Lambda ^{\perp } \leq \F_3^{20}$ such that 
$$D^{\perp } = \pi_t(\ker(\pi _f)) \leq \Lambda = \Lambda ^{\perp } \leq D .$$
Clearly also $d(\Lambda ) \geq d(D) \geq 6$, so $\Lambda $ is an extremal
ternary code of length 20.
There are 6 such codes, none of them has a proper overcode with minimum distance 6. 
\eb

\begin{rem}
If $\sigma \in \Mon (C) $ is some automorphism of order 4, then $\sigma ^2= -1 $ or
$\sigma ^2$ has Type $(24,0,0)$ in the notation of Proposition \ref{aut2}.
\end{rem}

\bew
Assume that $\sigma \in \Mon(C)$ has order 4 but $\sigma ^2 \neq -1$. 
Then $\tau = \sigma ^2$ is one of the automorphisms from Proposition \ref{aut2}
and so $\sigma $ is conjugate to some block diagonal matrix
$$\sigma \sim \diag( \left(\begin{array}{cccc} 0 & 1 & 0 & 0  \\ 0 & 0 & 1 & 0
\\ 0 & 0 & 0 & 1 \\ 1 & 0 & 0 & 0 \end{array} \right) ^{t/2} ,
\left(\begin{array}{cc} 0 & 1 \\ 1 & 0 \end{array} \right) ^{f/2} , \left(\begin{array}{cc} 0 & -1 \\ 1 & 0 \end{array} \right)^{a/2} )$$
If $t=22$ and $f=2$ then 
The fixed code of $\sigma $ is a self-dual code in 
$\langle (1,1,1,1) \rangle ^{t/2} \operp \langle (1,1) \rangle ^{f/2 } $
and $C(\sigma )^* \leq \F_3^{t/2+f/2} $ is a self-dual code 
with respect to the form $(x,y):= \sum _{i=1}^{t/2} x_iy_i - \sum _{i=t/2+1}^{t/2+f/2} x_iy_i $ which implies that $t/2-f/2$ is a multiple of $4$, a contradiction.
%So $t=24$ and the code $C(\sigma ) \cong \pi_t(C(\sigma ))  \leq \F_3^{12}$  is some 
%self-dual code of minimum weight $\geq 15/4$. 
%Since $d(\pi_t(C(\sigma )))$ is a multiple of 3, we get $d(\pi_t(C(\sigma ))) = 6$ and 
%$D:=\pi_t(C(\sigma )) $ is the extended quadratic residue code. 
%Moreover $C(\sigma )$ is obtained from $D$ by replacing $1$ by $(1,1,1,1)$ and 
%$0$ by $(0,0,0,0)$. 
%
\eb

For the two known extremal codes all automorphisms $\sigma $ of order 4 satisfy 
$\sigma ^2=-1$. It would be nice to have some argument to exclude the other
possibility.

%\subsection{Automorphisms of order 3.} 
%
%These automorphisms are treated with different methods in a different paper.
%We can show that an automorphism of order 3 has either no or 3 fixed points. 

\end{document}